\documentclass[12pt]{aipproc}
\usepackage{epsfig}

\newcommand{\de}{\Delta E}
\newcommand{\mbc}{M_{\rm bc}}

\newcommand{\bbar}{B\bar{B}}

\newcommand{\kstar}{\bar{K}^{*0}}
\newcommand{\kpi}{K^+\pi^-}

\newcommand{\kpipin}{\kpi\pi^0}
\newcommand{\kpipipi}{\kpi\pi^-\pi^+}
\newcommand{\phipi}{\phi\pi^+}

\newcommand{\kstark}{\kstar K^+}
\newcommand{\ksk}{K_S^0 K^+}
\newcommand{\ds}{D_s}
\newcommand{\dsp}{D_s^+}
\newcommand{\dsst}{D_s^*}

\newcommand{\dsj}{D_{sJ}}
\newcommand{\dsjp}{D_{sJ}^+}
\newcommand{\jp}{J/\psi}
\newcommand{\psip}{\psi(2S)}
\newcommand{\pipi}{\pi^+\pi^-}
\newcommand{\leplep}{l^+l^-}
\newcommand{\chic}{\chi_{c0}}
\newcommand{\eetodstdstb}{e^+e^- \to D^{(*)+}D^{(*)-}}
\newcommand{\eetodd}{e^+e^- \to D^+ D^-}
\newcommand{\eetoddst}{e^+e^- \to D^+ D^{*-}}
\newcommand{\eetodstdst}{e^+e^- \to D^{*+} D^{*-}}
\newcommand{\RM}{M_{\mathrm{recoil}}}
\newcommand{\RMD}{\Delta\RM}

\layoutstyle{6x9}

\begin{document}

\title{Recent results from Belle}

\author{P. Krokovny}{address={Budker Institute of Nuclear Physics, 
Novosibirsk, Russia}}

\begin{abstract}
New results on hadron physics from the Belle experiment are presented.
\vspace{1pc}
\end{abstract}

\maketitle

\section{Introduction}

These results are obtained using various data 
samples from 80~fb$^{-1}$ to 150~fb$^{-1}$ taken with
the Belle detector~\cite{NIM}. 
We identify $B$ candidates by two kinematic variables:
the energy difference, \mbox{$\de=(\sum_iE_i)-E_b$}, and the
beam constrained mass, $\mbc=\sqrt{E_b^2-(\sum_i\vec{p}_i)^2}$, where
$E_b=\sqrt{s}/2$ is the beam energy and $\vec{p}_i$ and $E_i$ are the 
momenta and energies of the decay products of the $B$ meson in the 
CM frame. 
The inclusion of charge conjugate modes is implicit
throughout this report.

\section{Observation of $0^+$ and $1^+$ broad $c\bar{u}$ states}

A study of charmed meson production in $B$ decays provides an
opportunity to test predictions of Heavy Quark Effective Theory (HQET)
and QCD sum rules. $B$ decays to $D^{(*)}\pi$ final states are its
dominant hadronic decay modes and are measured quite well~\cite{PDG}.
The large data sample accumulated in the Belle experiment allows to
study production of $D$ meson excited states.
$D^{**}$s are P-wave excitations of quark-antiquark  systems that
contain one charmed and one light ($u$,$d$) quark.

$B\to D^{**}\pi$ decays have been studied using the
$D^+\pi^-\pi^-$ and $D^{*+}\pi^-\pi^-$ final states~\cite{d2s_belle}.

\begin{figure}
  \includegraphics[width=0.33\textwidth] {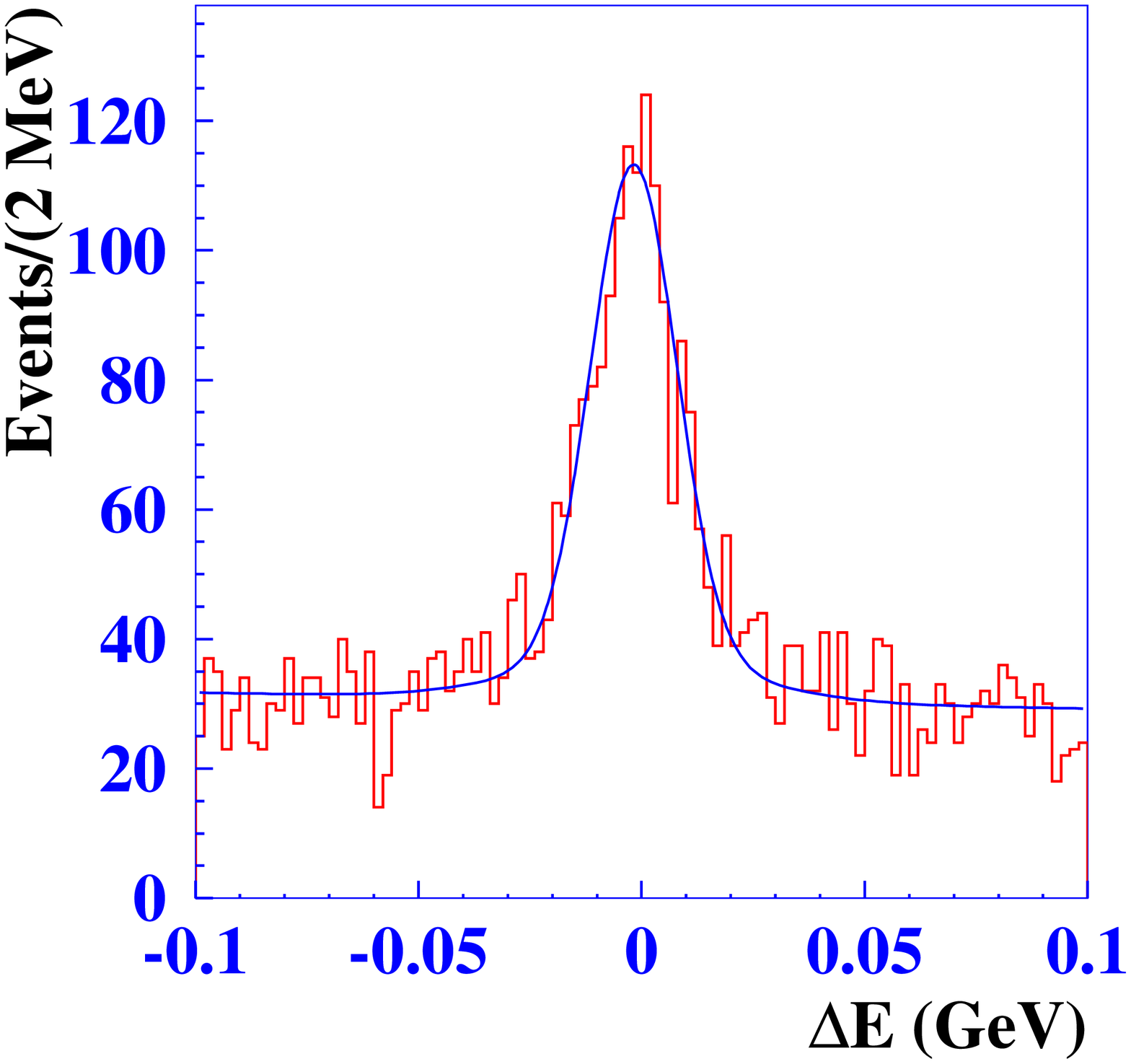}\hspace{1cm}
  \includegraphics[width=0.33\textwidth] {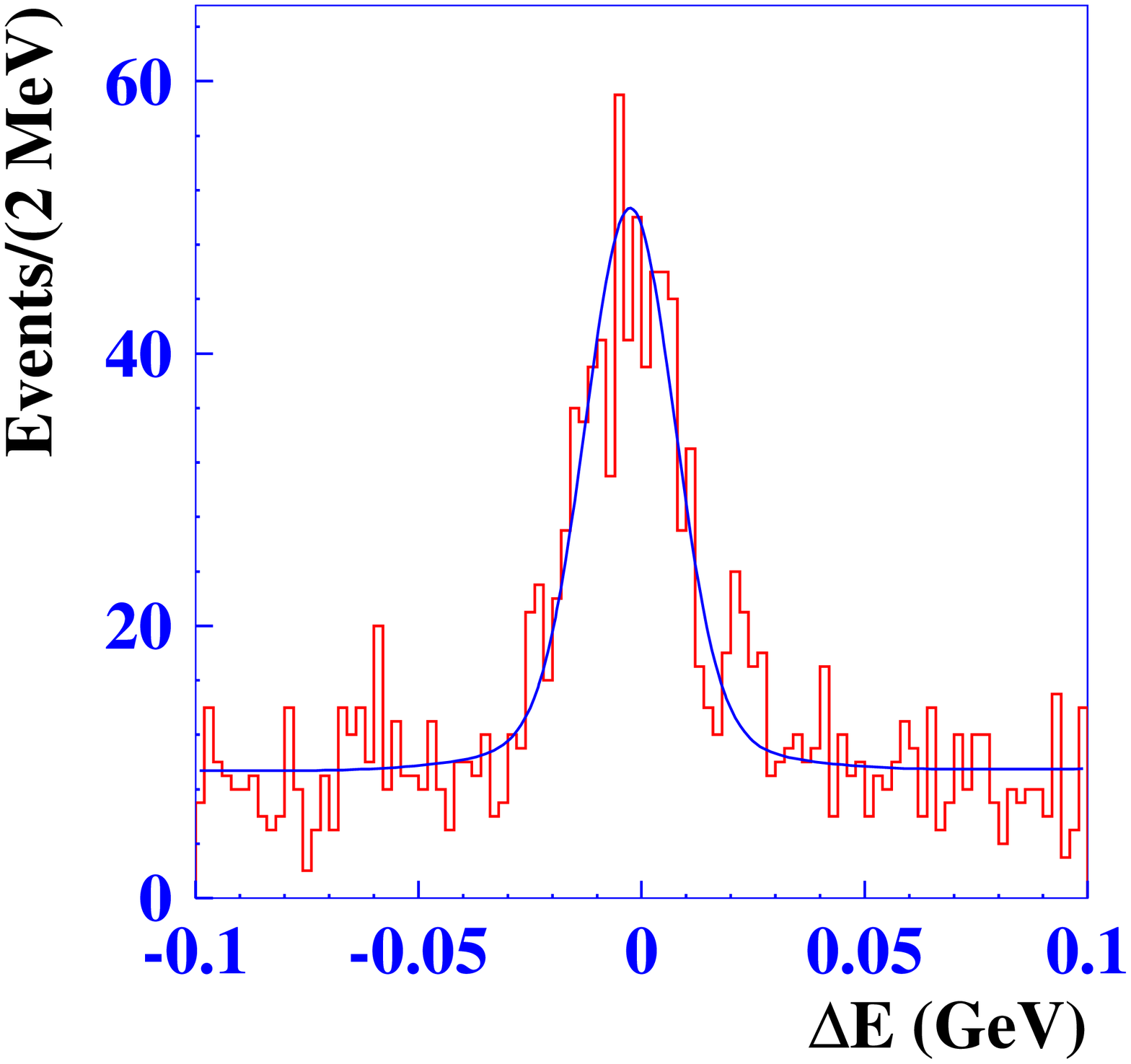}
\vspace*{-8mm}
  \caption{$\de$ distributions for the $B^-\to D^+\pi^-\pi^-$
  (left) and $B^-\to D^{*+}\pi^-\pi^-$ (right) candidates.}
  \label{d2s_de}
\end{figure}

Figure~\ref{d2s_de} shows the $\de$ distributions for the
$B^-\to D^+\pi^-\pi^-$ and $B^-\to D^{*+}\pi^-\pi^-$ candidates.
The following branching fractions are measured:
${\cal B}(B^-\to D^+\pi^-\pi^-)=(1.02\pm0.04\pm0.15)\times10^{-3}$ and
${\cal B}(B^-\to D^{*+}\pi^-\pi^-)=(1.25\pm0.08\pm0.22)\times10^{-3}$,
without any assumption about the intermediate final states.

\begin{table}
\footnotesize
\caption{Branching fractions and resonance parameters
for the $D^{(*)+}\pi^-\pi^-$ final states.}
\medskip
\label{d2s_res}
  \begin{tabular}{lrrr}\hline\hline
Mode  & \hspace*{-1cm}${\cal B}(B^-\to D_X[D^{(*)+}\pi^-]\pi^-)$, & 
$M(D_X)$, & $\Gamma(D_X)$,\\
& $10^{-4}$& MeV & MeV \\\hline
$B^-\to D_2^{*0}[D^+\pi^-]\pi^-$ & 
  $3.4\pm0.3\pm0.6\pm0.4$ & 
  $2462\pm2.1\pm0.5\pm3.3$ & $45.6\pm4.4\pm6.5\pm1.6$\\
$B^-\to D_0^{*0}[D^+\pi^-]\pi^-$ & 
  $6.1\pm0.6\pm0.9\pm1.6$ & 
  $2308\pm17\pm15\pm28$ & $276\pm21\pm18\pm60$\\
\hline
$B^-\to D_1^0[D^{*+}\pi^-]\pi^-$ & 
  $6.8\pm0.7\pm1.3\pm0.3$ & 
  $2421\pm1.5\pm0.4\pm0.8$ & $23.7\pm2.7\pm0.2\pm4.0$\\
$B^-\to D_2^{*0}[D^{*+}\pi^-]\pi^-$ & 
  $1.8\pm0.3\pm0.3\pm0.2$ & 
  \cite{d2s_note} & \cite{d2s_note}\\
$B^-\to D_1^{\prime 0}[D^{*+}\pi^-]\pi^-$ & 
  $5.0\pm0.4\pm1.0\pm0.4$ & 
  $2427\pm26\pm20\pm15$ & $384^{+107}_{-75}\pm24\pm70$\\
\hline\hline
  \end{tabular}
\end{table}

\begin{figure}
  \includegraphics[width=0.33\textwidth] {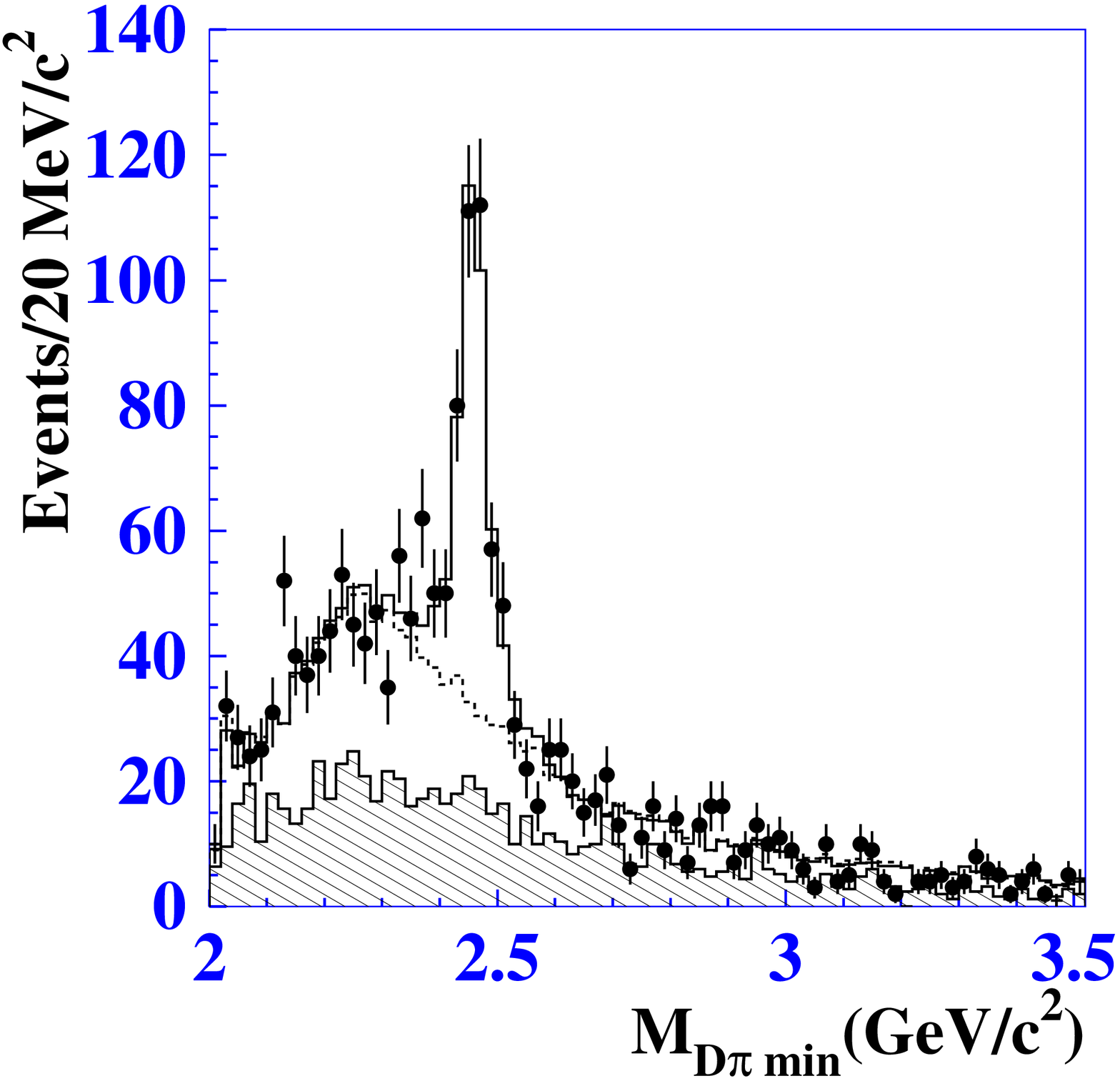}\hspace{1cm}
  \includegraphics[width=0.33\textwidth] {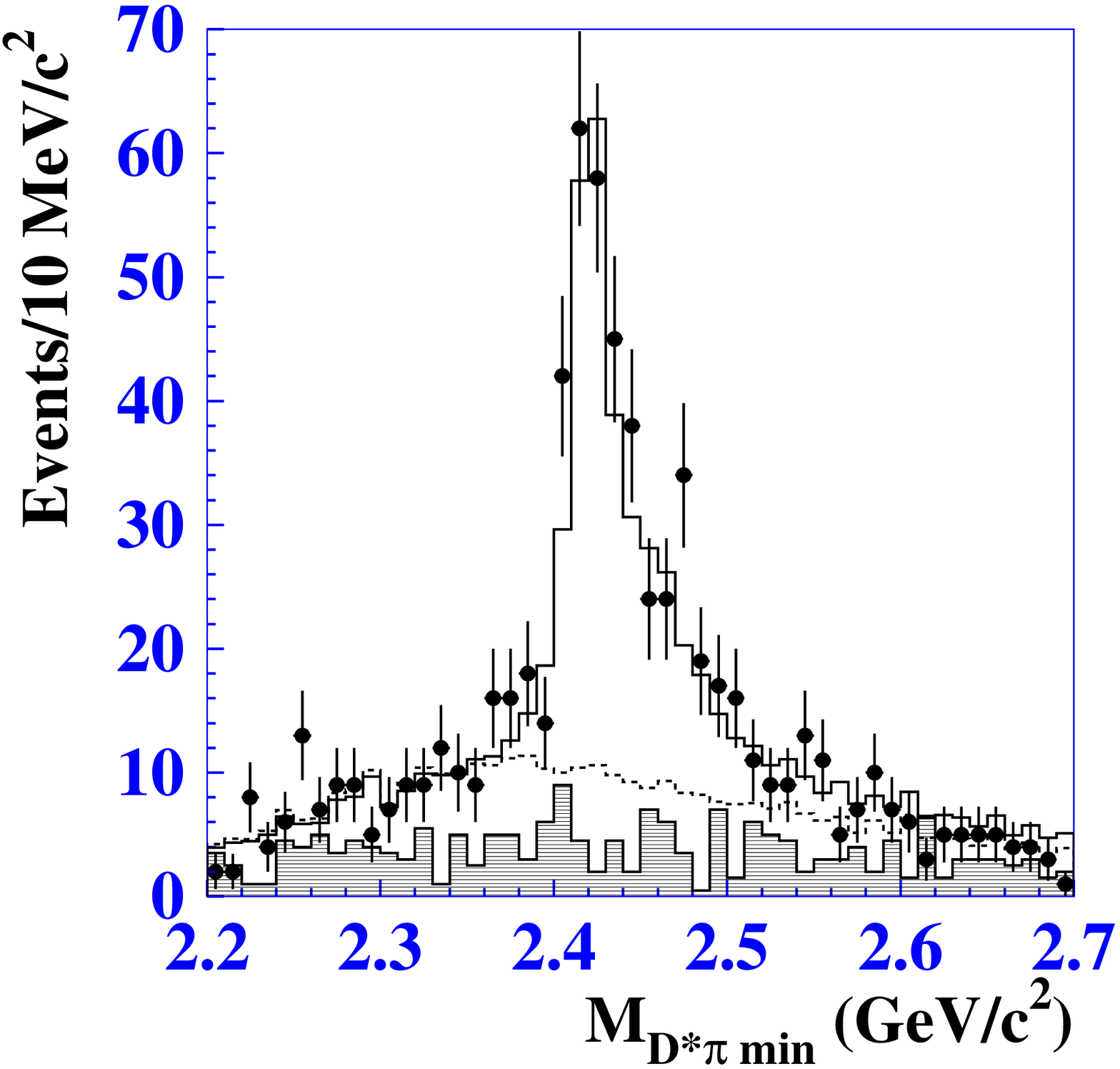}
\vspace*{-2mm}
  \caption{Minimal $D^+\pi^-$ (a) and $D^{*+}\pi^-$ (b) invariant
  mass distributions. The points with error bars correspond to
 the $B$ meson signal events,  the hatched histogram shows the sidebands.
 The open histogram is the result of a fit while the dashed one shows 
 the fit function without narrow resonance contribution.}
  \label{d2s_dalitz}
\end{figure}

To study the dynamics of $B\to D^{(*)}\pi\pi$ decays, analyses of
the Dalitz plots are performed.
The description of the Dalitz plot $D^+ \pi^- \pi^-$ includes 
amplitudes of the known $D_2^{*0}\pi^-$ mode, possible contributions
of the processes with virtual $D^{*0}\pi^-$ and 
$B^{*0}\pi^-$ production and an intermediate $D^+\pi^-$ broad 
resonance structure with free mass and width. 
Figure~\ref{d2s_dalitz}(a) shows the
$D^+\pi^-$  invariant mass distribution together with the 
resulting fit. A clear signal of the broad
resonance with $J^P=0^+$ is observed which can be identify as
the scalar $D_0^{*0}$ state.
The results of the mass, width and branching fraction products 
are presented in Table~\ref{d2s_res}.

For the $D^{*+}\pi^-\pi^-$ final state the fit of the 
density distribution is performed in four dimensional phase space
to take into account  the angles of the pion from $D^*$ decay. The fit 
function includes both known $D^{*0}_2$, $D^0_1$ intermediate state 
contributions and a broad $D^{*+}\pi^-$ resonance  with free
parameters. Figure~\ref{d2s_dalitz}(b) shows the
$D^{*+}\pi^-$  invariant mass distribution as well as the 
resulting fit. Together with the narrow resonances a clear signal 
of the broad state  is observed. The angular distribution of $D^*\pi$
from this state is consistent with $J^P=1^+$, $j_q=1/2$. 
This state can be identified as a P-wave
excitation of $c\bar{u}$ --  $D_1^{\prime0}$. The results of the mass, 
width and branching fraction products are presented in 
Table~\ref{d2s_res}. 

Together with observations of the broad resonances the 
branching ratios of B decay to the modes with known $D^{**}$:
$D_1^0\pi^-$ and $D_2^{*0}\pi^-$ have been measured.
Using these measurements the ratio of $D_2^{*0}$ branching
fractions $h={\cal B}(D_2^{*0}\to D^+\pi^-)/
{\cal B}(D_2^{*0}\to D^{*+}\pi^-)=1.9\pm 0.5$,
consistent with the world average $h=2.3\pm 0.6$~\cite{PDG}, is obtained.
The measured ratio $R={\cal B}(B^-\to D_2^{*0}\pi^-)/
{\cal B}(B^-\to D_1^0\pi^-)=0.77\pm 0.15$ is lower than the CLEO
measurement $1.8\pm 0.8$~\cite{r_cleo} (although the results are
consistent within errors) but is still a factor of two larger than the
factorization prediction~\cite{neubert}.
From our measurement it is impossible to determine  whether the
non-factorized part for  tensor and axial mesons is large, or whether
higher order corrections to the leading  factorized terms
should be taken into account.

Our measurements show that the narrow resonances
compose  $(36\pm 6)\%$ of the
$D\pi\pi$ decays and $(63\pm 6)\%$ of the $D^*\pi\pi$ decays.
This  result is inconsistent with the QCD sum rule prediction and may
indicate a large contribution from a color suppressed amplitude.

\section{Observation of new states $\dsjp(2317)$ and $\dsjp(2457)$}

The narrow $\ds \pi^0$ resonance at 2317 MeV, recently observed
by the BaBar collaboration~\cite{babar_dspi0}, is naturally
interpreted as a P-wave excitation of the $c\bar{s}$ system. The 
observation of a nearby  and narrow $\dsst \pi^0$ resonance by the 
CLEO collaboration~\cite{cleo_dspi0} supports this view, since the
mass difference of the two observed states is consistent with the 
expected hyperfine splitting for a P-wave doublet with total
light-quark angular momentum $j=1/2$~\cite{bardeen,pwave-doublets}.
The observed masses are, however, considerably lower than potential 
model predictions~\cite{bartelt}, and similar to those of the 
$c\bar{u}$ $j=1/2$ doublet states recently reported by 
Belle~\cite{d2s_belle}.
Measurements of the $\dsj$ quantum numbers and branching fractions 
(particularly those for radiative decays),
will play an important role in determining the nature of these states.

\begin{figure*}
\includegraphics[width=0.32\textwidth]{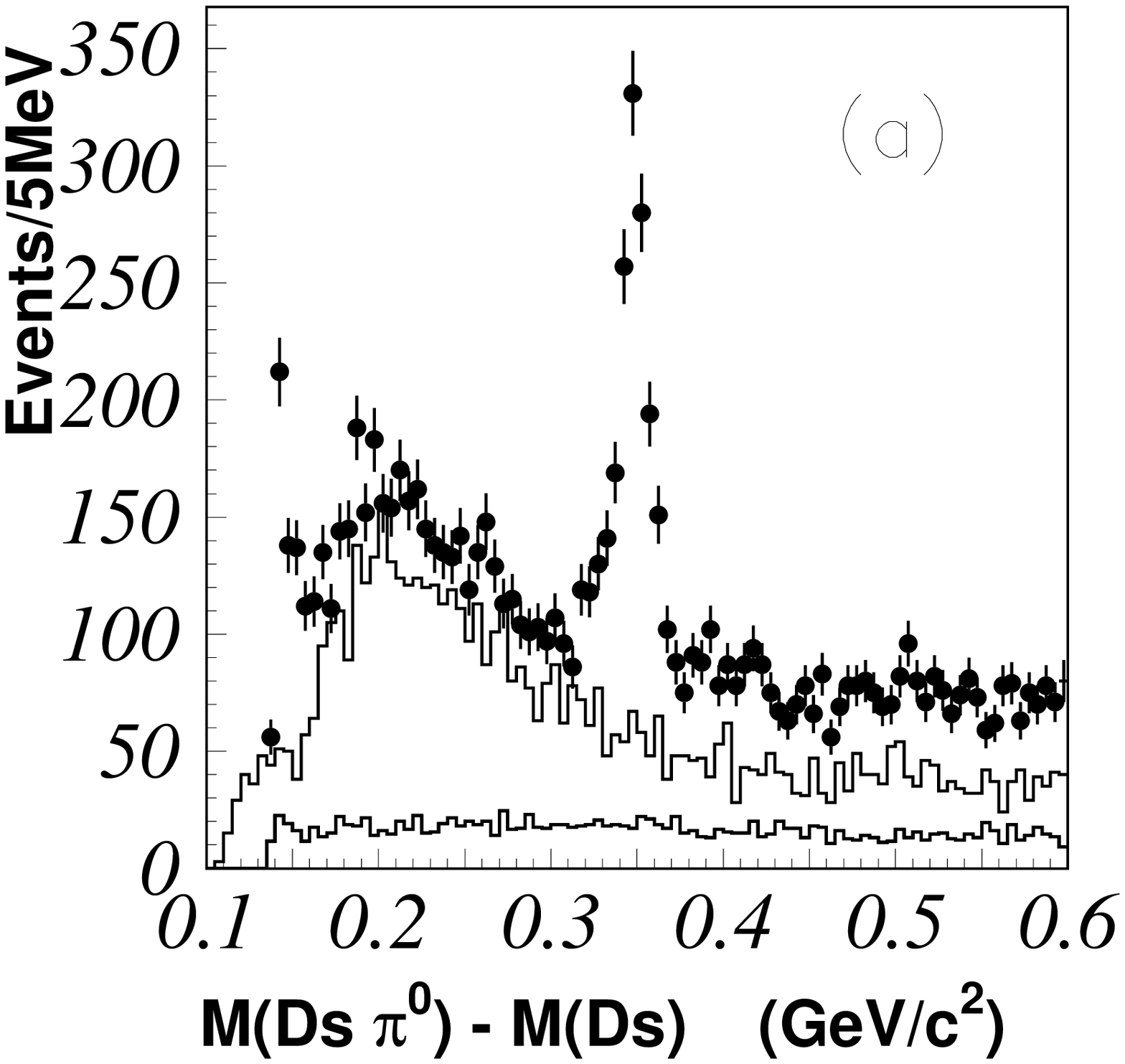}
\includegraphics[width=0.32\textwidth]{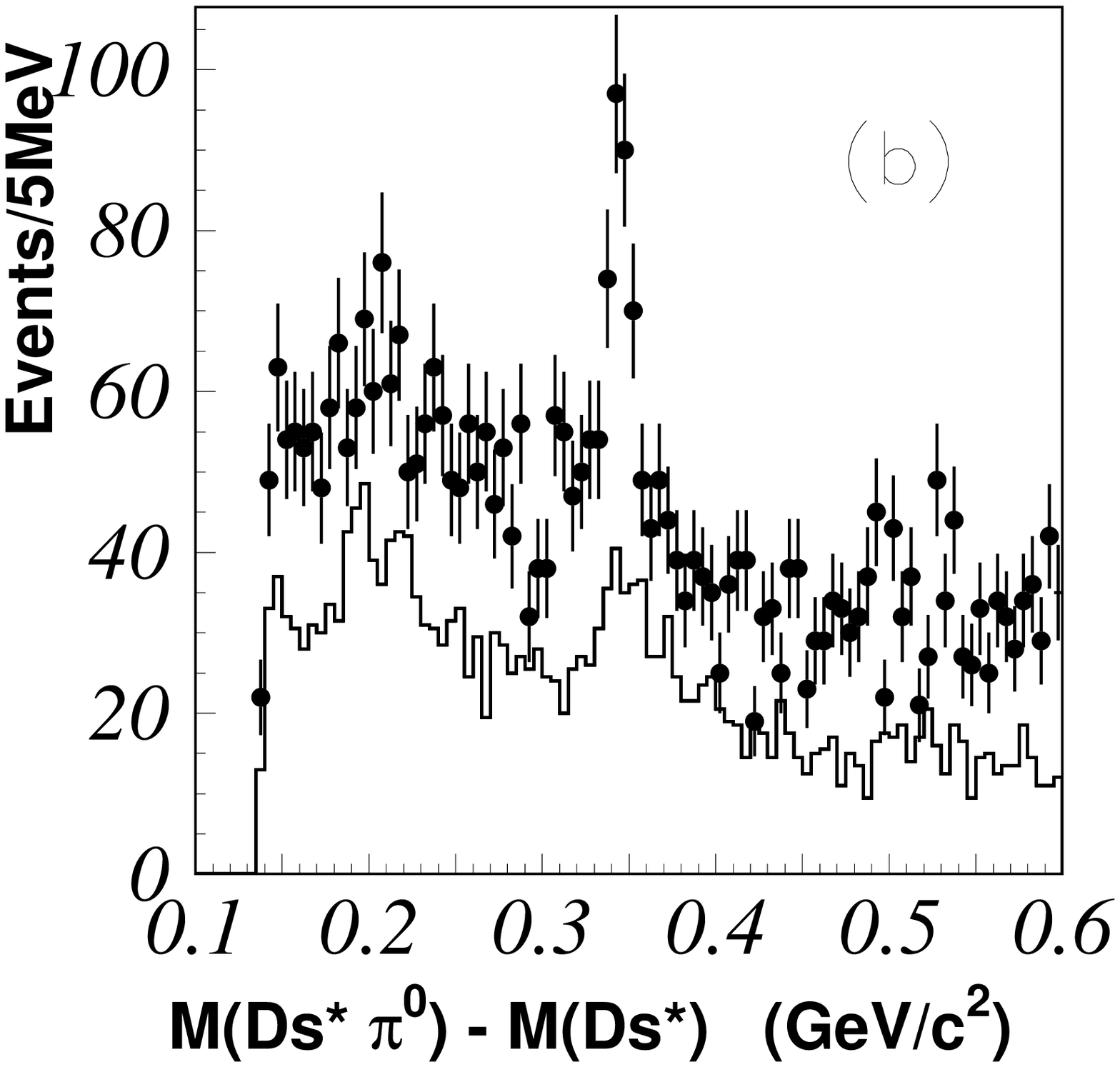}
\includegraphics[width=0.32\textwidth]{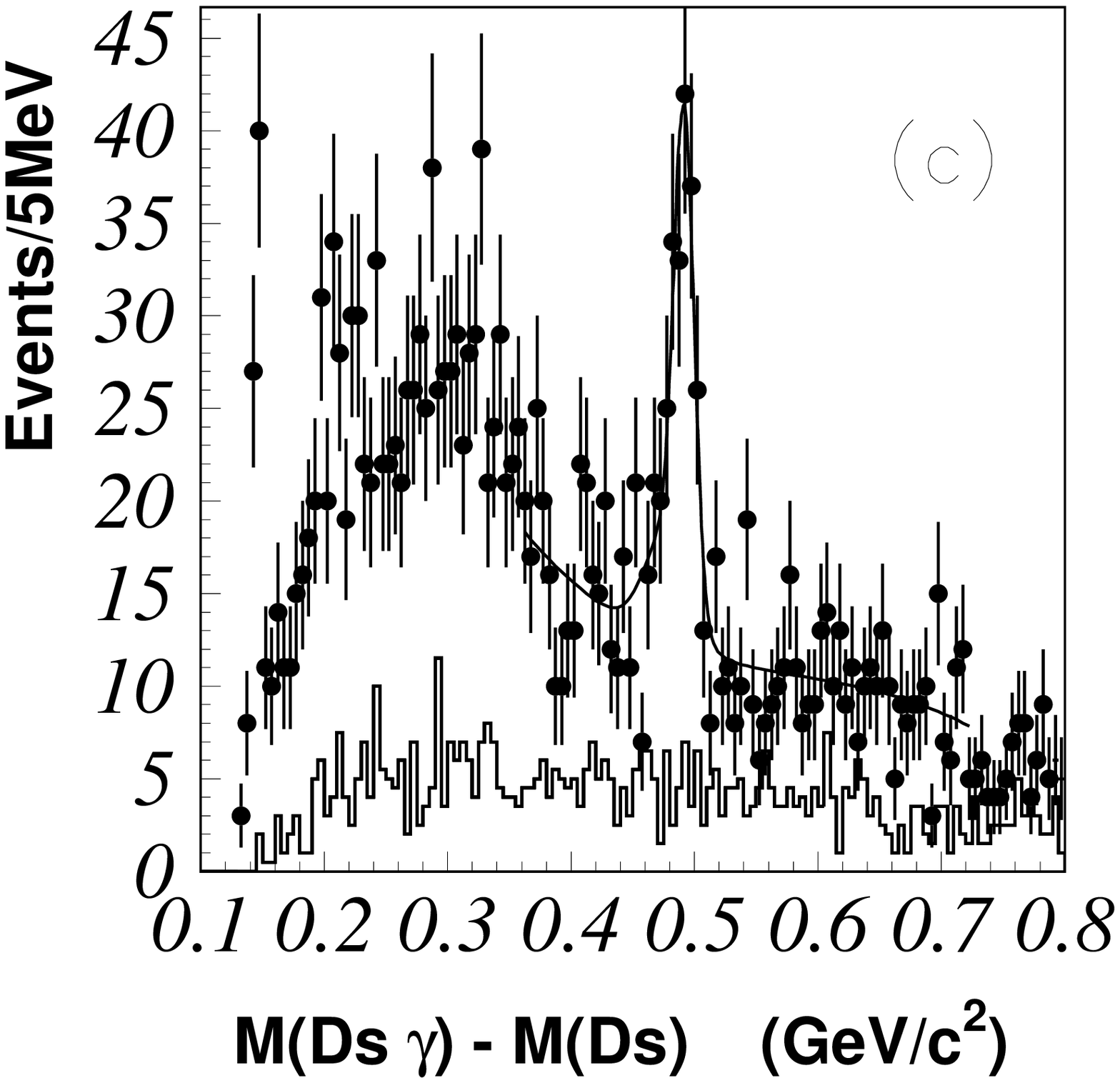}
\vspace*{-8mm}
\caption{$M(\ds\pi^0) - M_{\ds}$ (a),
$M(\dsst\pi^0) - M_{\dsst}$ (b) and 
$M(\ds\gamma) - M_{\ds}$ (c) mass-difference distributions.
The signal is described using a double Gaussian and a third-order 
polynomial for the background. 
The histogram shows no structure for the $D^{(*)+}_s$ sidebands.}
\label{dsj_cont}
\end{figure*}

We confirmed both resonances and measured masses for $0^+$ and
$1^+$ states to be $(2317.2\pm 0.5\pm 0.9)$~MeV and 
$(2456.5\pm 1.3\pm 1.3)$~MeV respectively~\cite{belle_dsj}. 
We also report the first observation of the radiative decay
$\dsj(2457)\to\ds\gamma$.
Figure~\ref{dsj_cont} shows the mass difference between the 
$D_s^{(*)}\pi^0$ and $D_s^{(*)}$ candidates.
The ratio $\frac{{\cal B}(\dsj(2457)\to\ds\gamma)}
{{\cal B}(\dsj(2457)\to\dsst\pi^0)}$ is found to be 
$0.55\pm 0.13\pm 0.08$.
     
\begin{figure}
  \includegraphics[width=0.4\textwidth] {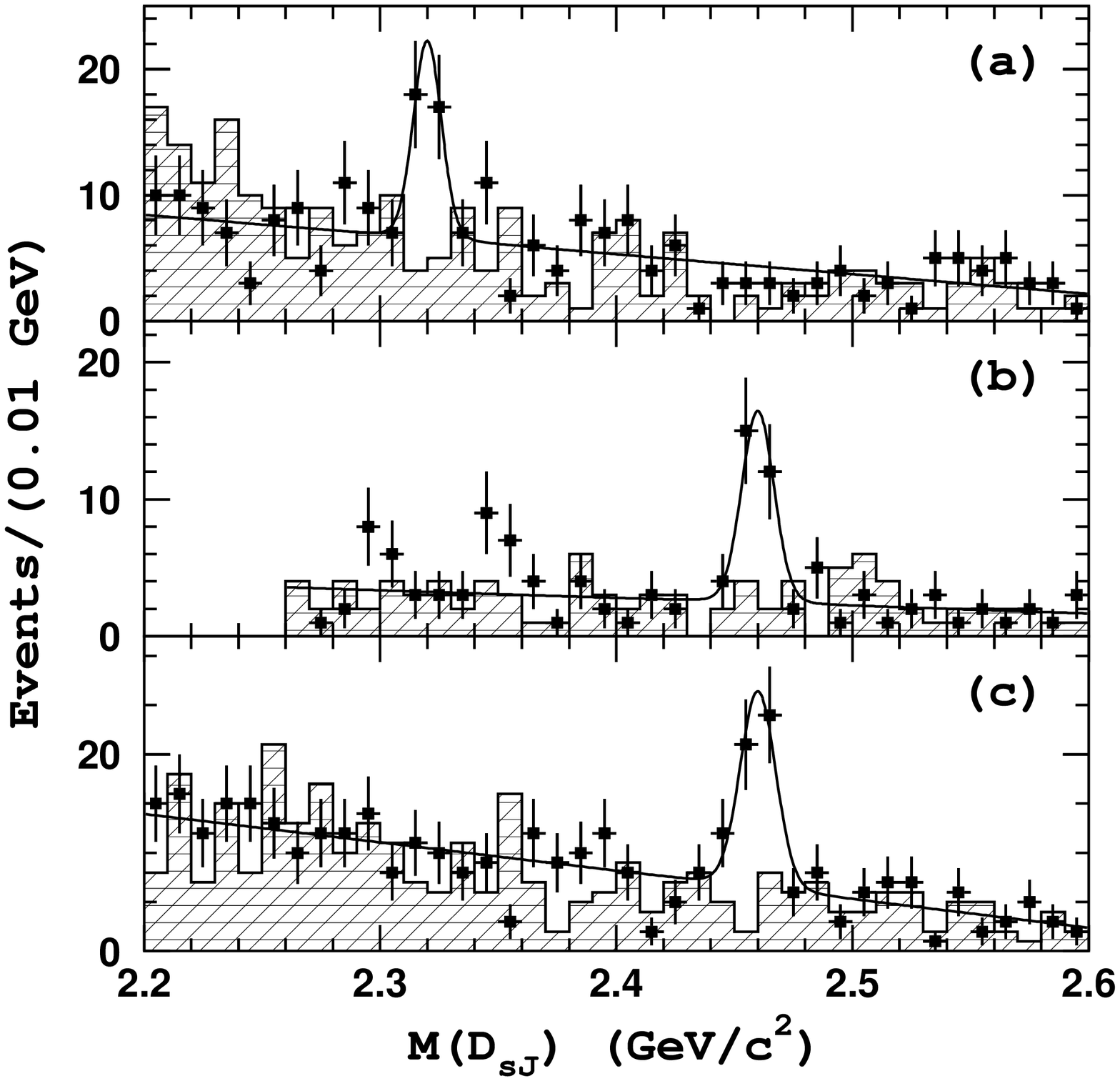}\hspace{1cm}
  \includegraphics[width=0.4\textwidth] {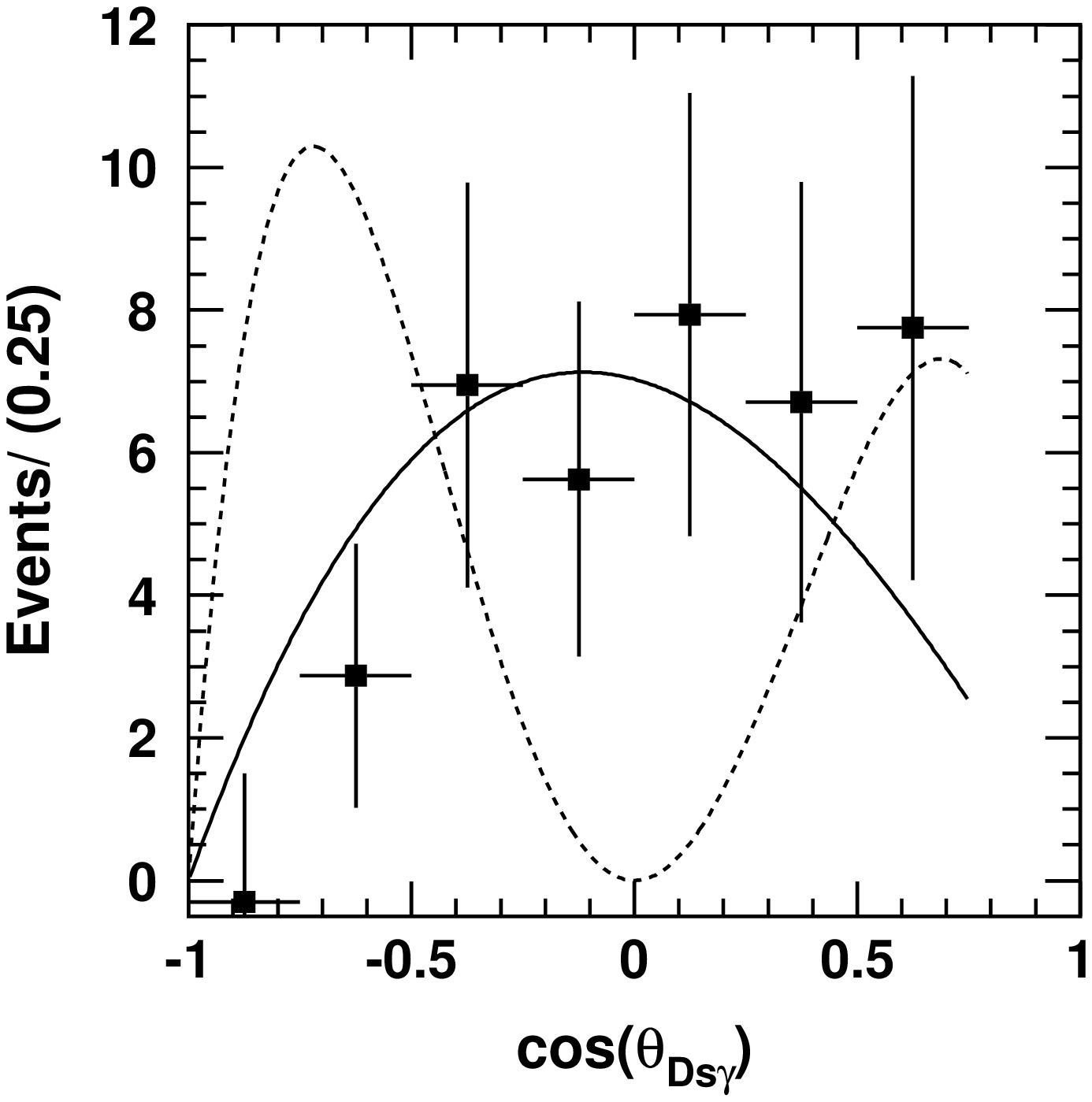}
\vspace*{-5mm}
  \caption{Left: $M(\dsj)$ distribution for the
    $B\to\bar{D}\dsj$ candidates: (a) $\dsj(2317)\to\ds\pi^0$, (b)
    $\dsj(2457)\to\dsst\pi^0$ and (c) $\dsj(2457)\to\ds\gamma$.
    Right: the $\dsj(2457)\to\ds\gamma$ helicity distribution.
    Points with errors represent the experimental data 
    and curves are the results of the fits.}
  \label{b2dsj_de}
\end{figure}

We also search for $\dsj$ production in 
$B\to D\dsj$ decays~\cite{dsj_bdecays}.
We reconstruct $\bar{D}^0 (D^-)$ mesons in the $\kpi$, $\kpipipi$ and
$\kpipin$ $(\kpi\pi^-)$ decay channels.
$\dsp$ mesons are reconstructed in the $\phipi$, $\kstark$ and $\ksk$
decay channels.
$\dsj$ candidates are reconstructed from $D_s^{(*)}$ mesons and
a $\pi^0$, $\gamma$, or $\pi^+\pi^-$ pair. The mass difference
$M(\dsj)-M(D_s^{(*)})$ is used to select $\dsj$ candidates.
We use central mass values of 2317~MeV and 2460~MeV
for $\dsj(2317)$ and $\dsj(2457)$ respectively and define signal
regions within 12~MeV for the corresponding mass difference.
We observe a clean signal for $B\to D\dsj(2317)[\ds\pi^0]$ and 
$B\to D\dsj(2457)[\dsst\pi^0]$. We also observe for the first time the
$\dsj(2457)\to\ds\gamma$ decay. Figure~\ref{b2dsj_de}(left)
shows the invariant mass distributions for these decays. The measured 
branching fractions are presented in Table~\ref{simulfit}.
We obtain the ratio $\frac{{\cal B}(\dsj(2457)\to\ds\gamma)}
{{\cal B}\dsj(2457)\to\dsst\pi^0)}=0.38\pm 0.11\pm 0.04$, which is
consistent with that from the continuum study. 

\begin{table}
\caption{$B\to D\dsj$ branching fractions.}
\medskip
\label{simulfit}
\small
\begin{tabular}{lcc}\hline\hline
Decay channel & ${\cal B}$, $10^{-4}$ & Signif.\\\hline
$B\to\bar{D} \dsj(2317)~[\ds\pi^0]$ &
$8.5^{+2.1}_{-1.9}\pm 2.6$ & $6.1\sigma$\\
$B\to\bar{D} \dsj(2317)~[\dsst\gamma]$ &
$2.5^{+2.0}_{-1.8}(<7.5)$ & $1.8\sigma$\\
$B\to\bar{D}\dsj(2457)~[\dsst\pi^0]$ &
$17.8^{+4.5}_{-3.9}\pm 5.3$ & $6.4\sigma$\\
$B\to\bar{D}\dsj(2457)~[\ds\gamma]$ &
$6.7^{+1.3}_{-1.2}\pm 2.0$ & $7.4\sigma$\\
$B\to\bar{D}\dsj(2457)~[\dsst\gamma]$ &
$2.7^{+1.8}_{-1.5}(<7.3)$ & $2.1\sigma$\\
$B\to\bar{D}\dsj(2457)~[\ds\pi^+\pi^-]$ & $<1.6$ & ---\\
$B\to\bar{D}\dsj(2457)~[\ds\pi^0]$ & $<1.8$ & ---\\\hline\hline
\end{tabular}
\end{table}

We also study the helicity distribution for the
$\dsj(2457)\to\ds\gamma$ decay.
The helicity angle $\theta_{\ds\gamma}$ is defined as the angle
between the $\dsj(2457)$ momentum in the $B$ meson rest frame and the
$\ds$ momentum in the $\dsj(2457)$ rest frame.
The $\theta_{\ds\gamma}$ distribution in the data 
(Fig.~\ref{b2dsj_de}(right)) is
consistent with MC expectations for the $J=1$ hypothesis for the
$\dsj(2457)$ ($\chi^2/$n.d.f$=5/6$), and contradicts the $J=2$ hypothesis
($\chi^2/$n.d.f.$=44/6$). The $J=0$ hypothesis is already ruled out by
the conservation of angular momentum and parity in
$\dsj(2457)\to\ds\gamma$.

\section{Observation of a new narrow
charmonium state in $B^{\pm}\to K^{\pm} \pipi\jp$ decay}

A major experimental issue for the $c\bar{c}$  charmonium particle
system is the existence of as yet unestablished charmonium states that
are expected to be below threshold for decays to open charm and, thus, 
narrow. These include the $n=1$ singlet P state, the $h_c(1{\rm P}),$ 
and possibly the $n=1$ singlet and triplet spin-2~D states, i.e. the 
$J^{PC}=2^{-+}~1^1{\rm D}_{c2}$ and $J^{PC}=2^{--}~1^3{\rm D}_{c2}$, 
all of which are narrow if their masses are below the $D\bar{D^*}$ 
threshold.  The observation of these states and the determination of
their masses would provide useful information about the spin 
dependence of the charmonium potential.

We report on an experimental study of the $\pipi\jp$
and $\gamma\chic$ mass spectra from exclusive $B^+\to K^+\pipi\jp$ and 
$K^+\gamma\chic$ decays~\cite{olsen} using a 152M $B\bar{B}$ event sample.
For the $B\to K \pi^+\pi^- J/\psi$ study we use events that
have a pair of well identified oppositely charged
electrons or muons with an invariant mass in the
range $3.077<M_{\leplep}<3.117$~GeV, a loosely identified
charged kaon and a pair of oppositely
charged pions.

\begin{figure}
\includegraphics[width=0.6\textwidth]{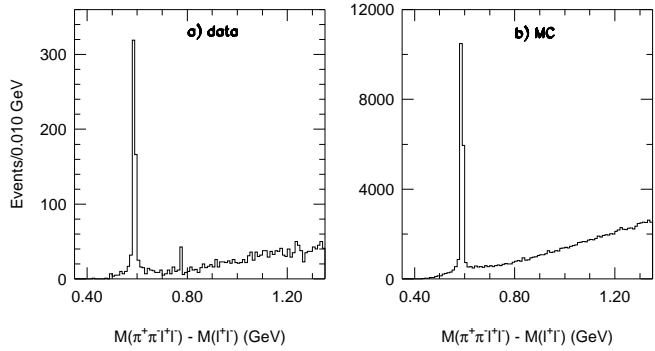}
\vspace*{-5mm}
\caption{Distributions of $M(\pi^+\pi^- \leplep)-M(\leplep)$ for
selected events in the $\de$-$\mbc$ signal region for
{\bf (a)} Belle data and  {\bf (b)} generic $\bbar$ MC
events .}
\label{fig:pipill-ll}
\end{figure}

Figure~\ref{fig:pipill-ll}(a) shows the distribution of
$\Delta M \equiv M(\pi^+\pi^- \leplep) - M(\leplep)$
for events in the $\de$-$\mbc$ signal region.
Here a large peak corresponding to $\psi(2S)\to\pi^+\pi^- J/\Psi$ is 
evident at 0.589~GeV. In addition, there is a significant spike in the 
distribution at 0.775~GeV.  Figure~\ref{fig:pipill-ll}(b) shows
the same distribution for a large sample of generic $\bbar$ Monte
Carlo (MC) events. Except for the prominent $\psi(2S)$ peak, the 
distribution is smooth and featureless.

\begin{table}
\caption{Results of the fits to the $\psip$ and $M=3872$~MeV
regions.  The errors are statistical only.}
\label{tb:fit_results}
\medskip
\begin{tabular}{lcc}\hline\hline
Quantity             & $\psip$ region  & $M=3872$~MeV region  \\
\hline
Signal events        &   $489\pm 23$   &   $35.7\pm 6.8$    \\
$M^{\rm meas}_{\pipi\jp}$ peak~&~~$3685.5\pm 0.2$~MeV &~~$3871.5\pm
0.6$~MeV\\
$\sigma_{M\pipi\jp}$ &   $3.3\pm 0.2$~MeV  &   $2.5\pm 0.5$~MeV  \\
\hline
\end{tabular}
\end{table}

\begin{figure}
\vspace*{3mm}
\includegraphics[angle=270,width=0.7\textwidth]{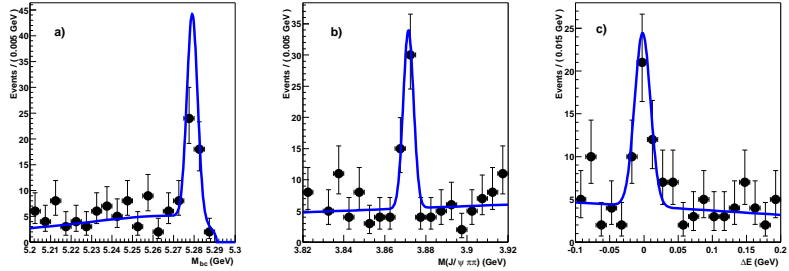}
\vspace*{-5mm}
\caption{ Signal-band projections of {\bf (a)} $\mbc$,
{\bf (b)} $M_{\pipi\jp}$ and {\bf (c)} $\de$ for the
$X(3872)\to\pipi\jp$ signal region with the results of the
unbinned fit superimposed.
}
\label{fig:X_pipijp_fit}
\end{figure}

We make separate fits to the data in the $\psi(2S)$
($3580~{\rm MeV} <M_{\pipi\jp}<3780$~MeV) and the $M=3872$~MeV
($3770~{\rm MeV} <M_{\pipi\jp}<3970$~MeV) regions using a
simultaneous unbinned maximum likelihood
fit to the $\mbc$, $\de$, and $M_{\pipi\jp}$ distributions.
The results of the fits are presented in Table~\ref{tb:fit_results}.
Figures~\ref{fig:X_pipijp_fit}(a), (b) and (c) show
the $\mbc$, $M_{\pipi\jp}$, and $\de$ signal-band projections for
the $M=3872$~MeV signal region, respectively.  The superimposed
curves indicate the results of the fit.
There are clear peaks with consistent yields
in all three quantities. The signal yield of $35.7\pm 6.8$~events
has a statistical significance
of $10.3\sigma$. In the following we refer to this as the $X(3872)$.

We determine the mass of the signal peak relative to the well measured
$\psi(2S) $ mass:
$M_X=M^{\rm meas}_X-M^{\rm meas}_{\psi(2S)}+M^{\rm PDG}_{\psi(2S)}=
3872.0 \pm 0.6 \pm 0.5~{\rm MeV}$.
Since we use the precisely known value of the $\psip$
mass~\cite{PDG} as a reference, the systematic error is small.
The measured width of the $X(3872)$ peak is
$\sigma = 2.5\pm 0.5$~MeV,
which is consistent with the MC-determined resolution and the value
obtained from the fit to the $\psip$ signal.
From this we infer a 90\% confidence level
(CL) upper limit of $\Gamma < 2.3$~MeV.

We determine a ratio of product branching fractions for
$B^+\to K^+ X(3872)[\pipi\jp]$ and $B^+\to K^+ \psip[\pipi\jp]$
to be $0.063 \pm 0.012 \pm 0.007$.

The decay of the $^3{\rm D}_{c2}$ charmonium state to $\gamma\chic$
is an allowed $E1$ transition with a partial width
that is expected to be substantially larger than that
for the $\pipi\jp$ final state; e.g.\ the authors
of Ref.~\cite{eichten} predict
$\Gamma(^3{\rm D}_{c2}\to\gamma\chic)
> 5\times \Gamma(^3{\rm D}_{c2}\to\pipi\jp)$.
Thus, a measurement of the width for this decay channel
can provide important information about the nature
of the observed state.  We searched for an $X(3872)$
signal in the $\gamma\chic$ decay channel, concentrating
on the $\chic\to\gamma\jp$ final state.

\begin{figure}
\vspace*{5mm}
\includegraphics[angle=270,width=0.7\textwidth]{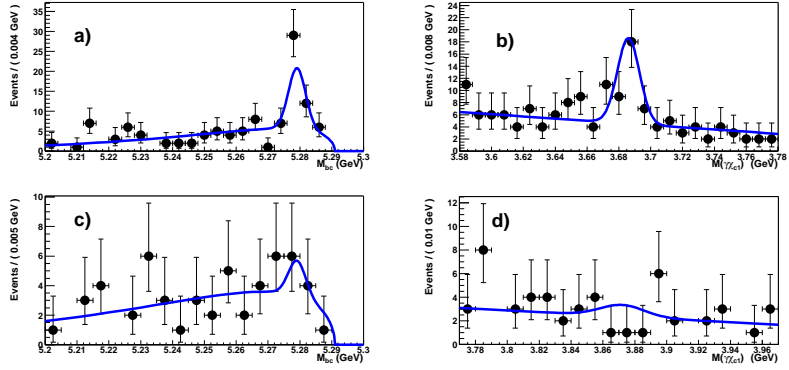}
\vspace*{-5mm}
\caption{ Signal-band projections of {\bf (a)} $\mbc$ and
{\bf (b)} $M_{\gamma\chic}$ for the
$\psip$ region with the results of the
unbinned fit superimposed. {\bf (c)} and {\bf (d)}
are the corresponding  results for
the $M=3872$~MeV region.}
\label{fig:gamchic1_fit}
\end{figure}

We select events with the same $\jp\to\leplep$
and charged kaon requirements
plus two photons, each with energy more than 40~MeV.
The signal-band projections of
$\mbc$ and $M_{\gamma\chic}$ for the $\psip$
region are shown in Figs.~\ref{fig:gamchic1_fit} (a)
and (b), respectively, together with curves that represent the results
of the fit.  The fitted signal yield is $34.1\pm 6.9 \pm 4.1$ events.
The number of observed events is
consistent with the expected yield of $26\pm 4$ events
based on the known $B\to K\psip$ and $\psip\to\gamma\chic$
branching fractions~\cite{PDG} and the MC-determined acceptance.

The results of the application of the same procedure to the 
$M=3872$~MeV region are shown in Figs.~\ref{fig:gamchic1_fit}(c) and (d).
Here, no signal is evident; the fitted signal yield
is $3.7\pm 3.7 \pm 2.2 $. From these results, we determine
a 90\% CL upper limit on the ratio of partial widths of
$\frac{\Gamma(X(3872)\to\gamma\chic)}{\Gamma(X(3872)\to\pipi\jp)}
<0.89$.
This limit on the $\gamma\chic$ decay width contradicts
expectations for the $^3{\rm D}_{c2}$ charmonium state.

The mass of the observed state is higher than potential model
expectations for the center-of-gravity ({\rm cog}) of the
$1^3{\rm D}_{cJ}$ states:
$M_{\rm cog}(1{\rm D})=3810$~MeV~\cite{tye,cornell}. 

In summary, we have observed a strong signal for a state
that decays to $\pipi\jp$ with
$M = 3872.0 \pm 0.6 \pm 0.5$~MeV and
$\Gamma < 2.3$~MeV (90\%~CL).
This mass value and the absence of a strong signal in
the $\gamma\chic$ decay channel are in some
disagreement with potential model
expectations for the $^3{\rm D}_{c2}$ charmonium state.
The mass is within errors of the $D^0\bar{D}^{*0}$ mass threshold
($3871.3 \pm 1.0$~MeV~\cite{PDG}), which is suggestive of
a loosely bound $D\bar{D}^*$ multiquark ``molecular state,''
as proposed by some authors~\cite{voloshin}.

\section{Measurement of the $\eetodstdstb$ cross-sections}
The processes $\eetodstdstb$
have not previously been observed at energies $\sqrt{s} \gg 2M_{D}$.
A calculation in the HQET approach based on the heavy-quark spin 
symmetry~\cite{grozin}, predicts cross-sections of about 
5~pb$^{-1}$ for $e^+e^- \to D\bar{D}^*$ and 
$e^+e^- \to D^*_T\bar{D}^*_L$ at $\sqrt{s} \sim 10.6$~GeV
(the subscripts indicate transverse [T] and longitudinal
[L] polarization of the $D^*$); the cross-section for $e^{+}e^{-} \to
D \bar{D}$ is expected to be suppressed by a factor of $\sim
10^{-3}$.

This analysis~\cite{uglov} is based on 88.9~fb$^{-1}$ of data 
taken at or near the $\Upsilon(4S)$ resonance. 
We reconstruct $D^0$ and $D^+$ mesons in the 
decay modes $D^0 \to K^- \pi^+$, $D^0 \to K^- \pi^+ \pi^+ \pi^-$ and 
$D^+ \to K^- \pi^+\pi^+$. $D^{*+}$ mesons are
reconstructed in the $D^0 \pi^+$ decay mode.

The processes $\eetodstdstb$ can be identified by
energy-momentum balance in fully reconstructed events that contain
only a pair of charm mesons.  However, the reconstruction efficiency 
is small in this case. Taking into account two body
kinematics, it is sufficient to reconstruct only one of the two
charmed mesons in the event to identify the processes of interest. 
We choose the mass of the system recoiling against the reconstructed 
$D^{(*)}$ ($\RM(D^{(*)+})$) as a discriminating variable:
$\RM(D^{(*)+})=\sqrt{(\sqrt{s}-E_{D^{(*)+}})^2-\vec{p}_{D^{(*)+}}^2}$,
where $E_{D^{(*)+}}$ and $\vec{p}_{D^{(*)+}}$ are the CM energy and
momentum of the reconstructed $D^{(*)+}$. For the signal a
peak in the $\RM$ distribution around the nominal mass of the recoiling
$D^-$ or $D^{*-}$ is expected. This method provides a significantly
higher efficiency, but also a higher background, in comparison to
full event reconstruction. For the $e^+ e^- \to D^{(*)+} D^{*-}$
processes we reconstruct in addition a slow pion from the 
$D^{*-}\to\bar{D}^0 \pi^-_{slow}$ decay. This reduces the 
background to a negligible level.

We calculate the difference between the masses of the systems
recoil mass against a $D^{(*)+} \pi^-_{slow}$ combination, and against
the $D^{(*)+}$ alone,
$\RMD\equiv \RM(D^{(*)+} - \RM(D^{(*)+} \pi^-_{slow}) )$.
The variable $\RMD$ peaks around the nominal $D^{*+}-D^0$ mass
difference with a resolution of $\sigma_{\RMD}\sim
1\,\mathrm{MeV}$ as found by Monte Carlo simulation.  For
$e^+ e^- \to D^{(*)+} D^{*-}$ we combine $D^{(*)+}$ candidates
together with $\pi^-_{slow}$ and require $\RMD$ to be within a $\pm
2\,\mathrm{MeV}$ interval around the nominal $D^{*+}-D^0$
mass difference.

\begin{figure*}
\centering
\includegraphics[width=0.33\textwidth]{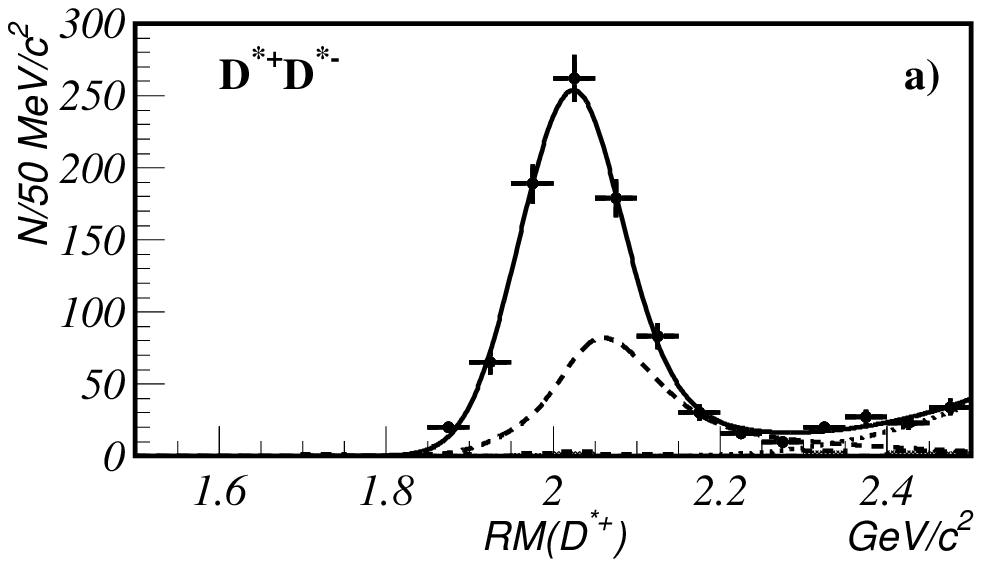}
\includegraphics[width=0.33\textwidth]{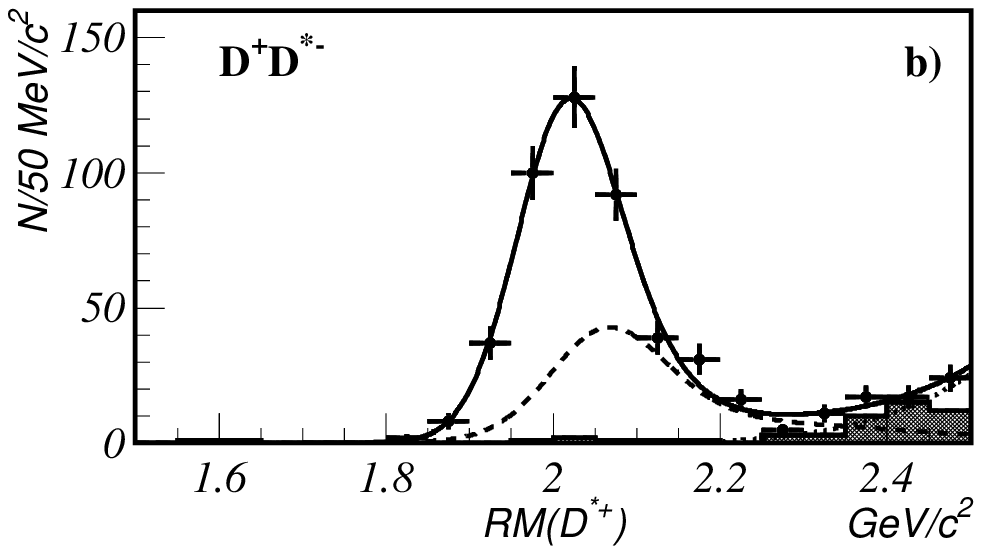}
\includegraphics[width=0.33\textwidth]{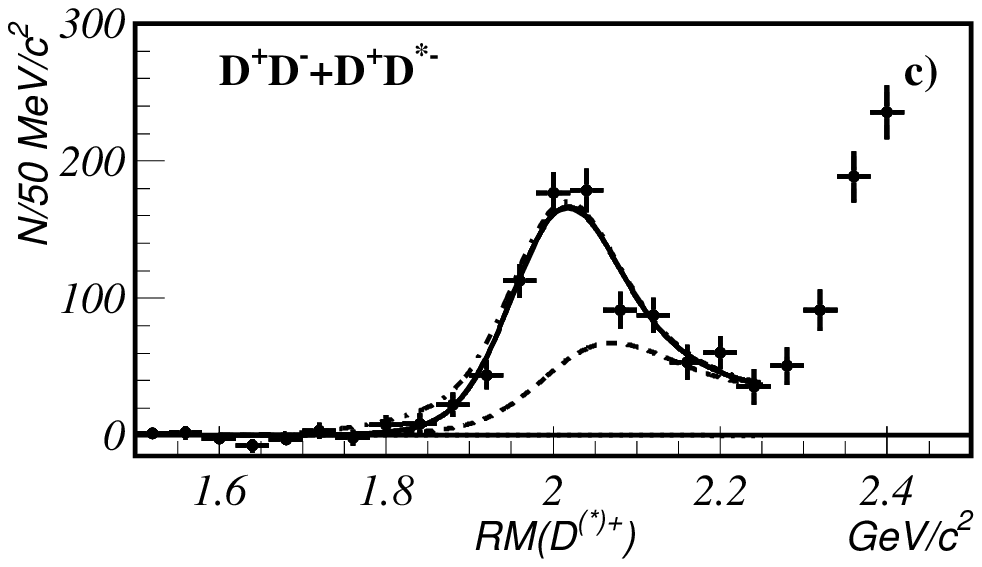}
\caption{Distributions of the mass of the system recoiling
against a) $D^{*+}$, and b) $D^+$.  Points with error bars show the
signal $\RMD$ region; hatched histograms correspond to $\RMD$
sidebands. The solid lines represent the fits described in the text;
the dashed lines show the contribution due to events with ISR photons
of significant energy.  The dotted lines show the expected background
contribution. c) The distribution of $\RM(D^+)$ without any
requirement on $\RMD$. }
\label{dstdst}
\end{figure*}

The $\RM(D^{*+})$ and $\RM(D^+)$ distributions are
shown in Figs.~\ref{dstdst}(a) and \ref{dstdst}(b), respectively. Clear 
signals are observed in both cases. The higher recoil mass tails in
the signal distribution are due to initial state
radiation (ISR). The hatched histograms show the $\RM$ distributions
for events in the $\RMD$ sidebands.

\begin{figure}
\centering
\includegraphics[width=0.33\textwidth]{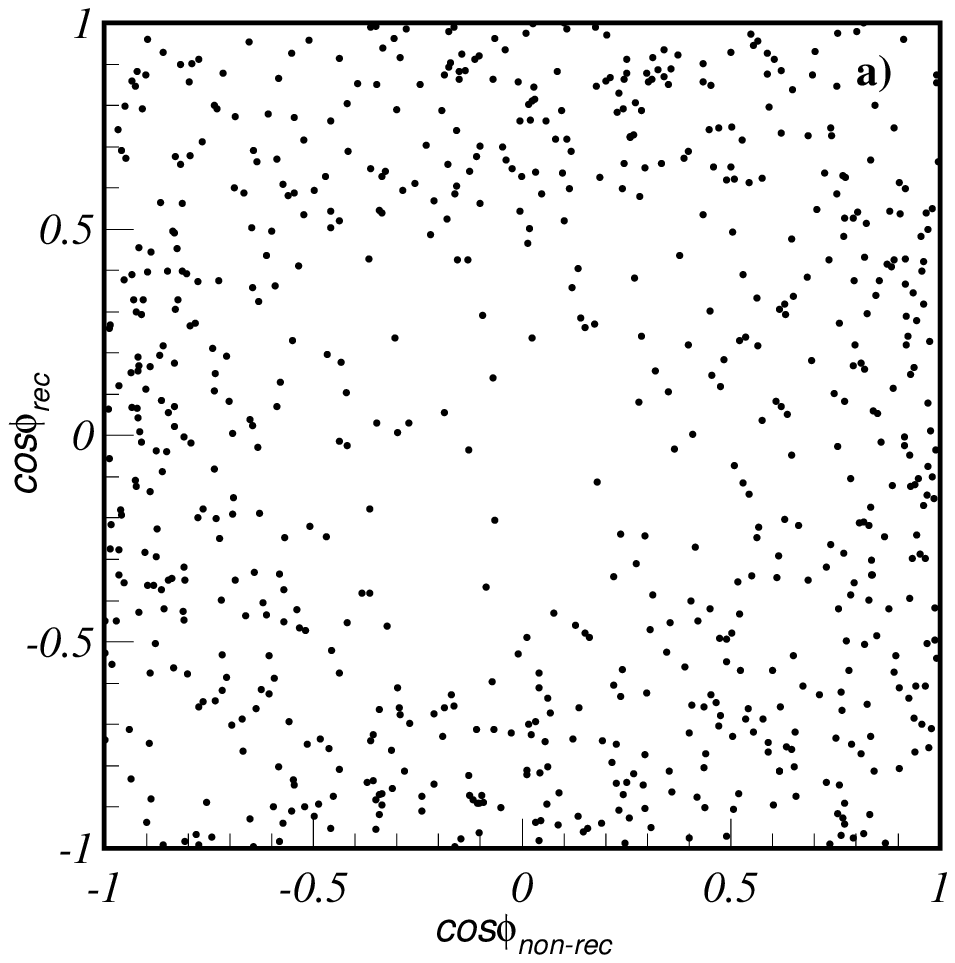}\hspace{1cm}
\includegraphics[width=0.4\textwidth]{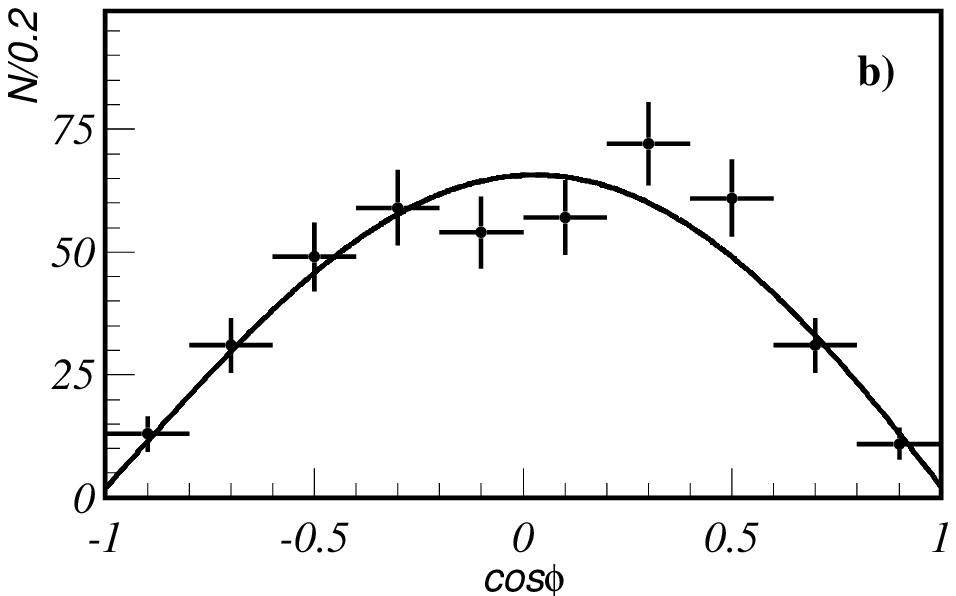}
\caption{a) The scatter plot $\cos(\phi_{D^*_{rec}})$ {\emph {vs}}
$\cos(\phi_{D^*_{non-rec}})$ ($\eetodstdst$). 
b) $D^{*+}$ meson helicity angle distribution for 
($\eetoddst$) signal candidates.}
\label{heli}
\end{figure}

Since the reconstruction efficiency depends on the production and
$D^{*\pm}$ helicity angles, we perform angular analysis before
computing  cross-sections. 
A scatter plot of the helicity angles for the two
$D^{*}$-mesons from $\eetodstdst$ ($\cos\phi(D^*_{rec})$ 
\emph {vs} $\cos\phi(D^*_{non-rec})$) for the recoil mass region 
$\RM(D^{*+})< 2.1$~GeV is shown in Fig.~\ref{heli}(a). The
distribution is fitted by a sum of three functions corresponding to
the $D^*_T D^*_T$, $D^*_T D^*_L$ and $D^*_L D^*_L$ final states,
obtained from Monte Carlo simulation. The fit finds the fractions of
$D^*_T D^*_T$, $D^*_T D^*_L$ and $D^*_L D^*_L$ final states to be
$(1.5\pm 3.6)\%$, $(97.2 \pm 4.8)\%$ and $(1.3 \pm 4.7)\%$,
respectively.  Figure~\ref{heli}(b) shows the $D^{*-}$ meson
helicity distribution for $\eetoddst$. The fraction of the 
$D^+D^{*-}_L$ final state is found from the fit to be equal to 
$(95.8 \pm 5.6)\%$.

We search for the process $\eetodd$ by studying the recoiling against
the reconstructed $D^+$ $(M_{recoil})$.
Fig.~\ref{dstdst}(c) shows the distribution of $\RM(D^+)$
after $D^+$ mass sideband subtraction. To extract the $\eetodd$ and 
$\eetoddst$ yields we fit this distribution with the sum of two signal 
functions corresponding to $D^-$ and $D^{*-}$ peaks and a threshold
function describing background events.
The fit finds $-13\pm 24$ events in the $D^-$ peak and $935\pm 42$ in
the $D^{*-}$ peak.
We obtains a $\eetoddst$ cross-section of $0.61\pm0.05$~pb
which agrees with the result using the $\RMD$ method.
For the $\eetodd$ cross-section we set an upper
limit of 0.04~pb at the 90\% confidence level.

In summary, we report the first measurement of the cross-sections for
the $\eetodstdst$ and $\eetoddst$ processes at 
$\sqrt{s}=10.6$~GeV to be $0.65 \pm 0.04 \pm 0.07$~pb and 
$0.71 \pm 0.05 \pm 0.09$~pb, respectively, and set an upper limit on
the $\eetodd$ cross-section of 0.04~pb at 90\% CL.
The measured cross-sections are an order of
magnitude lower than those predicted in Ref.~\cite{grozin}, but
their relative sizes are as predicted: the cross-sections for
$\eetodstdst$ and $\eetoddst$
are found to be close each other, while
the cross-section for $\eetodd$ is much smaller. The helicity
decomposition for $\eetodstdst$ is found to be saturated
by the
$D^{*\pm}_T D^{*\mp}_L$ final state (the fraction is equal to $(97.2
\pm 4.8) \%$) and for $\eetoddst$ --- by the $D^*_L$ final 
state ($95.8\pm5.6\%$), in good agreement with the predictions of
Ref.~\cite{grozin}.

\section{Observation of $\eta_c(2S)$ production and its mass measurement}
Belle recently observed  the $\eta_c(2S)$ production in exclusive $B$ 
decays to $KKK^0_{S}\pi$, where the $\eta_c(2S)$ is reconstructed in
the $K^{\pm}K^0_{S}\pi^{\mp}$ final state. 
The mass was measured to be 
$(3654\pm 6\pm 8)$~MeV~\cite{olsen_etac2s}  
which is much larger than the previous Crystal Ball measurement of 
$(3594\pm 5)$~MeV~\cite{crball_etac2s}.  
This year Belle also observed $\eta_c(2S)$ production 
($108\pm 24$ events) in double charmonia events 
$e^+e^-\to J/\psi \eta_c(2S)$ and confirmed a higher $\eta_c(2S)$ 
mass~\cite{belle_conf_0331_pasha}. 

\section{Conclusion}
We have observed a strong signal for a new charmonium state
that decays to $\pipi\jp$ with
$M = 3872.0 \pm 0.6 \pm 0.5$~MeV, $\Gamma<2.3$~MeV at 90\%~CL.
We confirm the observation of $\dsj(2317)$ and $\dsj(2457)$ and report
the first observation of the decay $\dsj(2457)\to\ds\gamma$. We also
observe $\dsj$ production in $B$ decays.
In $B^-\to D^{(*)+}\pi^-\pi^-$ decays all four P-wave $D^{**}$ have
been observed and their parameters have been measured.
For the broad $D_0^{*0}$ and $D_1^{*0}$ states
there are the first measurements.


\begin{thebibliography}{99}
\bibitem{pwave-doublets} {In the heavy $c$-quark approximation, one
    expects two doublets of $c\bar{s}$ states with quantum numbers 
    $J^P=0^+$, $1^+$ ($j = 1/2$) and $J^P=1^+$, $2^+$ ($j = 3/2$).
    The second doublet has already been observed in $DK$ and $D^*K$ 
    decays.}
\end{thebibliography}
\end{document}